\documentclass[conference]{IEEEtran}
\IEEEoverridecommandlockouts
\usepackage{cite}
\usepackage{amsmath,amssymb,amsfonts}

\usepackage{algorithmic}
\usepackage{graphicx}
\usepackage{textcomp}
\usepackage{xcolor}
\usepackage{times}
\usepackage{latexsym}

\usepackage[T1]{fontenc}

\usepackage[utf8]{inputenc}

\usepackage{microtype}

\usepackage{inconsolata}

\usepackage{graphicx}
\usepackage{amsmath}
\usepackage{subcaption}
\usepackage{multirow}
\usepackage{booktabs}
\usepackage{graphicx}
\usepackage{amsmath}
\usepackage{subcaption}
\usepackage{multirow}
\def\BibTeX{{\rm B\kern-.05em{\sc i\kern-.025em b}\kern-.08em
    T\kern-.1667em\lower.7ex\hbox{E}\kern-.125emX}}
\begin{document}

\title{LRCTI: A Large Language Model-Based Framework for Multi-Step Evidence Retrieval and Reasoning in Cyber Threat Intelligence Credibility Verification\\
{\footnotesize \textsuperscript{*}}
}

\author{\IEEEauthorblockN{1\textsuperscript{nd} Fengxiao Tang}
\IEEEauthorblockA{\textit{Central South University}\\
Changsha, China \\
tangfengxiao@csu.edu.cn}
\and
\IEEEauthorblockN{2\textsuperscript{nd} Huan Li}
\IEEEauthorblockA{\textit{Central South University}\\
Changsha, China \\
lihuan199911@gmail.com}
\and
\IEEEauthorblockN{3\textsuperscript{rd} Ming Zhao}
\IEEEauthorblockA{\textit{Central South University}\\
Changsha, China \\
meanzhao@csu.edu.cn}
\and
\IEEEauthorblockN{4\textsuperscript{rd} Zongzong Wu}
\IEEEauthorblockA{\textit{Central South University}\\
Changsha, China \\
wzy\_cookie@163.com}
\and
\IEEEauthorblockN{5\textsuperscript{rd} Shisong Peng}
\IEEEauthorblockA{\textit{Central South University}\\
Changsha, China \\
244703048@csu.edu.cn}
\and
\IEEEauthorblockN{6\textsuperscript{rd} Tao Yin}
\IEEEauthorblockA{\textit{Zhongguancun Laboratory}\\
Beijing, China \\
yintao@zgclab.edu.cn}
}

\maketitle

\begin{abstract}
Verifying the credibility of Cyber Threat Intelligence (CTI) is essential for reliable cybersecurity defense. However, traditional approaches typically treat this task as a static classification problem, relying on handcrafted features or isolated deep learning models. These methods often lack the robustness needed to handle incomplete, heterogeneous, or noisy intelligence, and they provide limited transparency in decision-making—factors that reduce their effectiveness in real-world threat environments. To address these limitations, we propose LRCTI, a Large Language Model (LLM)-based framework designed for multi-step CTI credibility verification. The framework first employs a text summarization module to distill complex intelligence reports into concise and actionable threat claims. It then uses an adaptive multi-step evidence retrieval mechanism that iteratively identifies and refines supporting information from a CTI-specific corpus, guided by LLM feedback. Finally, a prompt-based Natural Language Inference (NLI) module is applied to evaluate the credibility of each claim while generating interpretable justifications for the classification outcome. Experiments conducted on two benchmark datasets, CTI-200 and PolitiFact show that LRCTI improves F1-Macro and F1-Micro scores by over 5\%, reaching 90.9\% and 93.6\%, respectively, compared to state-of-the-art baselines. These results demonstrate that LRCTI effectively addresses the core limitations of prior methods, offering a scalable, accurate, and explainable solution for automated CTI credibility verification.
\end{abstract}


\section{Introduction}
As cyberattacks grow increasingly complex and adaptive, verifying the credibility of Cyber Threat Intelligence (CTI) has emerged as a critical challenge in cybersecurity research. The accuracy and timeliness of CTI directly influence the effectiveness of defense strategies and incident response systems \cite{1}. However, current methods for CTI credibility verification primarily relies on rule-based heuristics, statistical models, or supervised deep learning techniques \cite{3}. While these approaches have shown promising results in large-scale detection tasks, they often struggle with key limitations in real-world threat environments—particularly in terms of evidence completeness, interpretability, and generalizability.

\begin{figure}[t]
\centering
\includegraphics[height=0.24\textheight,width=0.9\columnwidth]{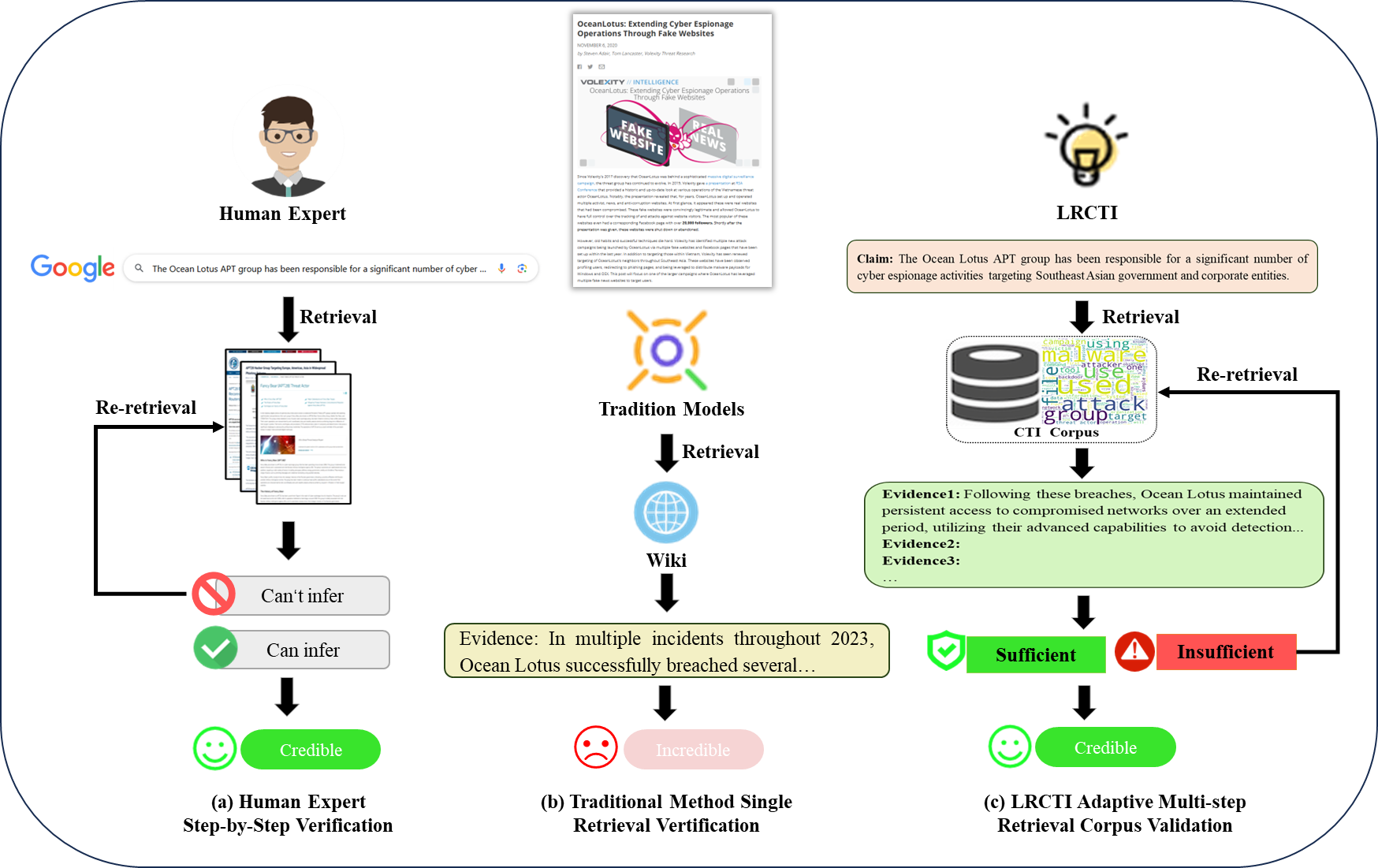}
\caption{Comparison of Different CTI Credibility Verification Methods: (a) Human Expert Verification: Information is manually collected through search engines and verified step by step, relying heavily on analyst experience. This process is time-consuming and potentially inconsistent. (b) Traditional LLM-Based Single-Step Retrieval: Relies on one-shot retrieval over general-purpose corpora, often resulting in fragmented or semantically mismatched evidence, leading to unreliable credibility judgments. (c) The Proposed LRCTI Framework: Performs adaptive multi-step retrieval over a dedicated CTI corpus, combined with structured reasoning to generate evidence-rich, interpretable, and highly accurate verification results.}
\label{fig1}
\end{figure}

Traditional algorithms typically function as “black boxes,” making it difficult for analysts to understand how a given decision was made. Furthermore, the diversity and inconsistency of CTI sources—ranging from technical reports and threat feeds to open-source and dark web content—introduce additional uncertainty, increasing the risk of both false positives and false negatives. Conceptually, CTI credibility verification shares similarities with fake news detection, where validating the authenticity of information requires contextual evidence. Techniques from that domain, including content-based retrieval \cite{4} and evidence-grounded reasoning \cite{5, 6, 7, 8, 9}, demonstrate value in controlled settings but often fail to generalize to the dynamic, domain-specific nature of cyber threat intelligence.

Recent advances in Large Language Models (LLMs) \cite{12} offer new opportunities for flexible, evidence-driven reasoning. LLMs have shown strong capabilities in natural language understanding, summarization, and multi-step inference. However, applying LLMs to CTI remains challenging due to the unstructured, noisy, and fragmented nature of real-world intelligence. Traditional approaches that process entire reports tend to incorporate irrelevant or redundant information, leading to inefficient retrieval and unreliable outputs.

To address these limitations, we propose LRCTI, a novel LLM-based framework for CTI credibility verification. LRCTI emulates the reasoning workflow of human analysts by decomposing the task into three stages: extracting focused claims, retrieving multi-step evidence from a CTI-specific corpus, and generating interpretable decisions. Specifically, LRCTI first applies a claim summarization module to condense verbose CTI reports into concise, verifiable propositions. It then performs iterative evidence retrieval, guided by model feedback, to build context-rich support for each claim. Finally, it uses prompt-based natural language inference to determine credibility while providing human-readable justifications.

As illustrated in Figure \ref{fig1}, LRCTI offers a significant departure from traditional methods. Human analysts rely on manual web search and step-by-step verification, which is labor-intensive and potentially inconsistent across analysts and contexts. Conventional one-shot LLM systems use shallow retrieval over general corpora (e.g., Wikipedia \cite{2}), often producing fragmented or irrelevant evidence. In contrast, LRCTI leverages domain-specific multi-step retrieval and structured inference, resulting in more robust, contextualized, and explainable outputs.

This paper represents the first application of multi-step LLM-driven reasoning to CTI credibility verification. Our key contributions are as follows:
\begin{itemize}
 \item We propose LRCTI, a novel framework that integrates LLMs with multi-step retrieval and reasoning for scalable, interpretable CTI credibility verification.
\item Multi-Step Evidence Retrieval Module: This module adopts an adaptive multi-step retrieval strategy that dynamically associates relevant evidence, reduces manual intervention, and significantly enhances the interpretability and reliability of the verification results.
\item Prompt-Based Natural Language Inference Module: Utilizing LLM-guided prompt templates, this module generates structured credibility decisions along with human-readable justifications, improving transparency and fostering user trust.
\item Experimental Validation and Performance: LRCTI is evaluated on two benchmark datasets, CTI-200 and PolitiFact, achieving F1-Macro and F1-Micro scores of 90.9\% and 93.6\%, respectively, and significantly outperforming state-of-the-art baselines in both accuracy and interpretability.
\end{itemize}

\section{Related Work}

\subsection{Credibility Verification in Cyber Threat Intelligence}
The credibility of CTI has long been identified as a fundamental barrier to its operational value and adoption \cite{13}. Existing literature highlights the inconsistencies in CTI collection, formatting, and analysis across platforms, which, compounded by the dynamic and fragmented nature of threats, makes credibility verification both difficult and underexplored. Most prior studies focus on source-level credibility rather than the content-level verifiability of intelligence. Commonly considered factors include timeliness, accuracy, completeness, relevance, and similarity \cite{15}.
To address these challenges, a number of automated evaluation models have been proposed. Andrew et al. \cite{17} proposed enriching IOCs through correlation across OSINT sources. Liu et al. \cite{18} used a DNN to evaluate content credibility based on quality metrics, but neglected time sensitivity and information completeness. Cheng et al. \cite{19} employed RNNs and node-path reliability algorithms but assumed static source credibility and freshness, which may not hold in practice. Li et al. \cite{20} adopted a DBN model incorporating content and temporal features, yet did not consider inter-document dependencies. Ermerins et al. \cite{21} analyzed cross-source overlaps to infer credibility, but ignored deeper semantic reasoning.

In contrast to prior work, we introduce LRCTI, which performs content-level credibility verification through multi-step retrieval and reasoning over a CTI-specific corpus. Rather than relying solely on static features or single-pass classification, our approach iteratively retrieves evidence, captures context dependencies, and generates interpretable conclusions with minimal human input.

\subsection{Retrieval-Augmented Language Models (RAG-LLMs)}

Retrieval-augmented generation (RAG) enhances LLM performance by supplementing model input with external knowledge, thereby mitigating hallucination, improving factual consistency, and increasing interpretability \cite{23}. RAG has demonstrated success in tasks such as question answering, Fact-Checking and Fake News Detection  \cite{27,28}. Recent RAG+LLM frameworks such as FLARE \cite{29}, RePlug \cite{30}, and ProgramFC \cite{31} focus on evidence selection from general corpora (e.g., Wikipedia) through either dense or sparse retrieval.
Our approach differs in two key aspects. First, LRCTI uses a CTI-domain-specific corpus, rather than open-domain sources, to ensure domain relevance and minimize noise. Second, LRCTI employs a multi-step, LLM-guided retrieval loop, incorporating feedback from the reasoning module to iteratively refine the evidence set, which enhances both retrievals precision and contextual alignment.

\subsection{Natural Language Inference}
Natural Language Inference (NLI) plays a central role in aligning claims with evidence by predicting logical entailment, contradiction, or neutrality. While traditional NLI tasks rely on pre-trained models, recent advances in LLM prompting strategies—such as Chain-of-Thought \cite{33}, ReAct \cite{24}, and ThinkingTrees \cite{34} —have significantly improved multi-step reasoning capabilities.
In the CTI setting, however, challenges such as input length constraints, evidence fragmentation, and reasoning clarity remain. While methods like cueing \cite{35}, instruction tuning \cite{36}, and RLHF \cite{37} improve alignment in general domains, they often lack robustness in handling domain-specific, multi-source CTI.
LRCTI extends LLM-based reasoning with prompt-driven NLI over retrieved evidence, enabling not only claim verification but also the generation of interpretable justifications. The design integrates retrieval and reasoning in a unified feedback loop, optimizing both factual grounding and user trust.

\section{Method}
\begin{figure*}[ht!]
\centering
\includegraphics[width=1.0\textwidth]{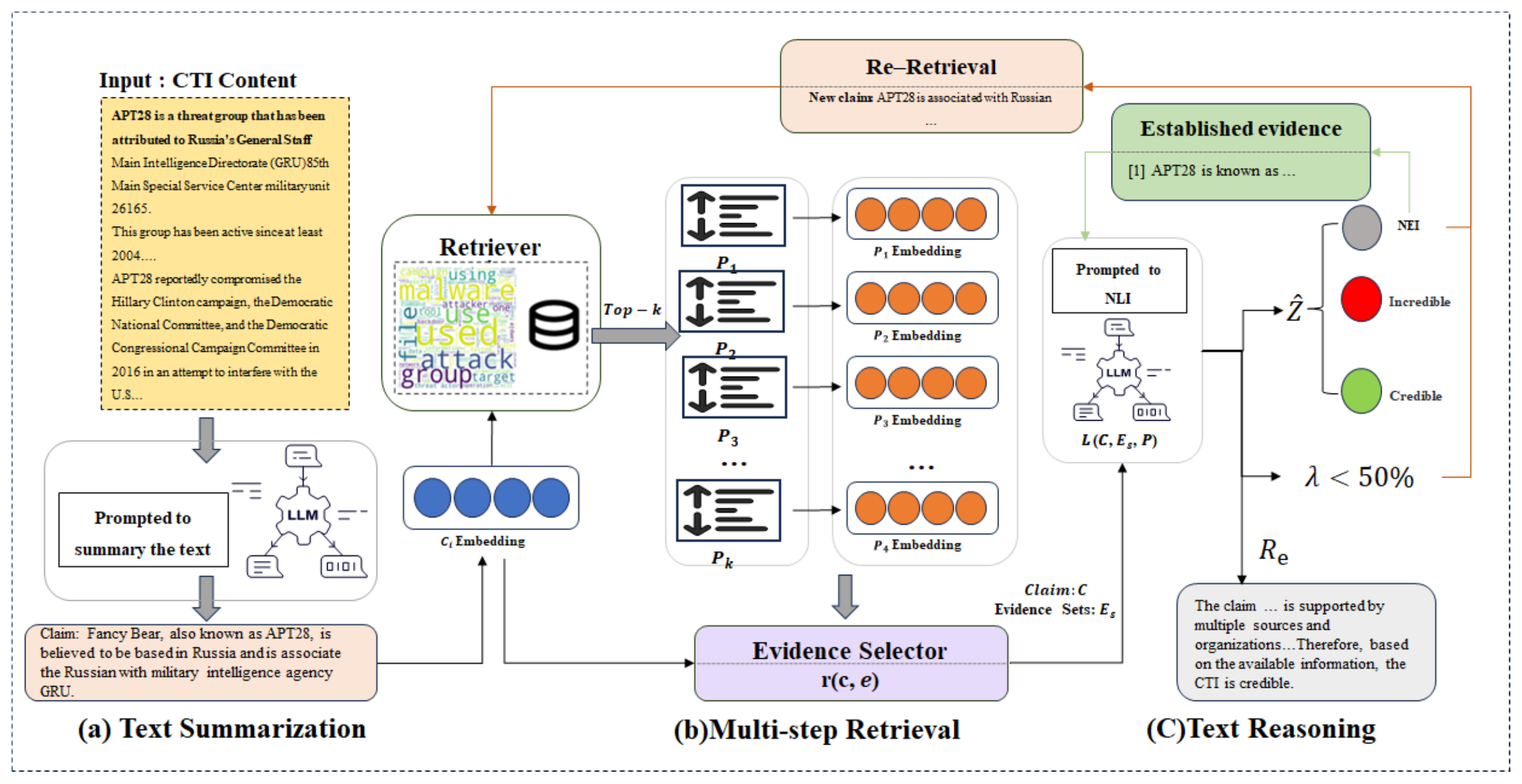} %
\caption{Our framework consists of three key steps: (a) \textbf{Text Summarization}: The input CTI report is condensed into a concise and verifiable claim using LLM-based prompt summarization.
(b) \textbf{Multi-Step Retrieval}: The Retriever identifies the top-$k$ relevant paragraphs, which are semantically embedded and filtered by the Evidence Selector based on correlation $r(c, e)$. If the evidence is insufficient or $\lambda < 50\%$, the system triggers re-retrieval and updates the established evidence pool and refined claims.
(c) \textbf{Text Reasoning}: The retrieved evidence set $E_s$ is passed to an LLM-based Natural Language Inference (NLI) module to determine a credibility judgment $\hat{z} \in \{\text{credible}, \text{incredible}, \text{NEI}\}$ along with an interpretable explanation $R_e$.
}
\label{fig2}
\end{figure*}
In this section, we present our proposed model, \textbf{LRCTI}, which performs CTI credibility verification using a structured, LLM-driven pipeline. Given a raw cyber threat intelligence report as input, the framework begins by applying a \textit{text summarization module} to extract concise and verifiable claims, denoted as $\mathbf{C}$. Subsequently, a set of relevant paragraphs $\mathbf{P = \{P_1, P_2, P_3, \ldots, P_k\}}$ is retrieved from a domain-specific CTI corpus. From these paragraphs, an evidence selection process identifies supporting fragments, forming an evidence set $\mathbf{E}_\mathbf{s} = \{\mathbf{E}_1, \mathbf{E}_2, \mathbf{E}_3, \ldots, \mathbf{E}_j\}$. A LLM then evaluates the relevance and sufficiency of the collected evidence. If the current evidence set is deemed adequate, the model proceeds to generate the final output. Otherwise, it initiates additional retrieval steps to refine and augment the evidence set. To ensure cost-efficiency and accessibility, OpenAI’s GPT-3.5-turbo is employed as the backbone LLM throughout the framework. The output of LRCTI includes a predicted credibility label $\hat{y} \in \{\text{credible}, \text{incredible}\}$ for the input claim, along with a generated explanation $R_e$, where $R_e = L(C, E_s)$. Here, $L$ denotes the reasoning function powered by the LLM. As shown in Figure~\ref{fig2}, the LRCTI framework consists of three main components: the text summarization module, the multi-step retrieval module, and the text reasoning module, which collectively form the core of the verification process.

\subsection{Text Summarization Module}

A summary is commonly defined as ``a text extracted from one or more documents that captures the most important information from the original content, typically within half its length.'' The goal of automatic text summarization is to produce concise, fluent, and informative summaries through machine automation—particularly useful in the context of CTI, where intelligence reports often contain substantial volumes of technical text. Cybersecurity analysts rely on accurate summaries to quickly extract critical content for credibility verification. Due to the high resource demands of fine-tuning models such as Pegasus \cite{38}—which requires large-scale training data and complex hyperparameter tuning—we adopt a prompt-based approach using LLM instead. Rather than retraining models, \textit{LLM prompt engineering} leverages the capabilities of powerful pre-trained models (e.g., BERT, GPT, T5) \cite{39,40,41} by designing targeted prompts that guide the model’s attention toward salient content during generation.

We begin by identifying a set of key sentences from the CTI document. Let the set of sentences be denoted as \(S = \{s_i\}_{i=1}^N\), where \(N\) is the total number of sentences. The importance of each sentence is quantified by a weighted score \(w(s_i)\), computed based on three factors: (i) the ROUGE1-F1 similarity between the sentence and the entire document, (ii) its positional weight \(p(s_i)\), and (iii) its semantic similarity with other sentences \(q(s_i)\). The final score is given by:

\begin{equation}
   w(s_i) = \text{ROUGE1-F1}(s_i, S) + \alpha \cdot p(s_i) + \beta \cdot q(s_i)
    \label{eq1}
\end{equation}

Where \(\alpha\) and \(\beta\) are hyperparameters controlling the contribution of each factor. The top \(m\) sentences with the highest scores are selected to form the subset \(R\), which serves as the input for summary generation.

We then apply prompt engineering to formulate a query \(P(R)\) that instructs the LLM to generate a summary based on the selected sentence set \(R\). 

The LLM generates the summary \(C\) using the prompt:

\begin{equation}
    C = \text{L}(P(R))
    \label{eq3}
\end{equation}

This approach eliminates the need for additional model pre-training or fine-tuning. By carefully crafting prompts, pre-trained LLM can be repurposed for high-quality summary generation with minimal overhead. It offers a flexible and efficient solution, particularly suited to scenarios where critical information must be extracted rapidly from large-scale, noisy, and technical CTI reports.

\subsection{Multi-Step Evidence Retrieval Module}

\textbf{Paragraph Retrieval}: The paragraph retrieval component focuses on selecting paragraphs relevant to a given claim \( C \) from the CTI domain corpus. Specifically, a set of top-\(k\) paragraphs \( P = \{p_1, p_2, \cdots, p_k\} \) is retrieved as:

\begin{equation}
P = R(C, \mathcal{P}, k) = \text{Top-}k_{p \in \mathcal{P}} \thinspace r(C, p)
\label{eq3}
\end{equation}

\begin{figure}[!h]
    \centering
    \begin{subfigure}[b]{0.95\columnwidth}
        \centering
        \includegraphics[width=\textwidth]{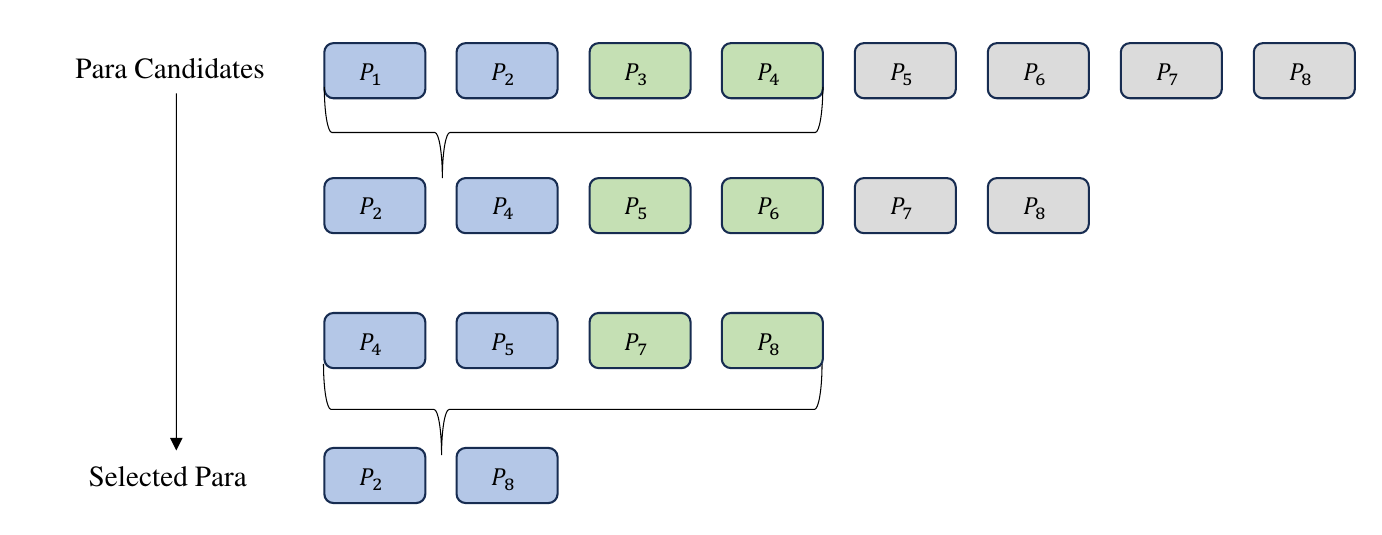}
        \caption{Progressive paragraph selection. Blue and green blocks represent current paragraphs \( P \) and new candidates \( P_c^\ast \), respectively.}
        \label{fig4a}
    \end{subfigure}
    
    \vspace{0.5cm}
    
    \begin{subfigure}[b]{0.95\columnwidth}
        \centering
        \includegraphics[width=\textwidth]{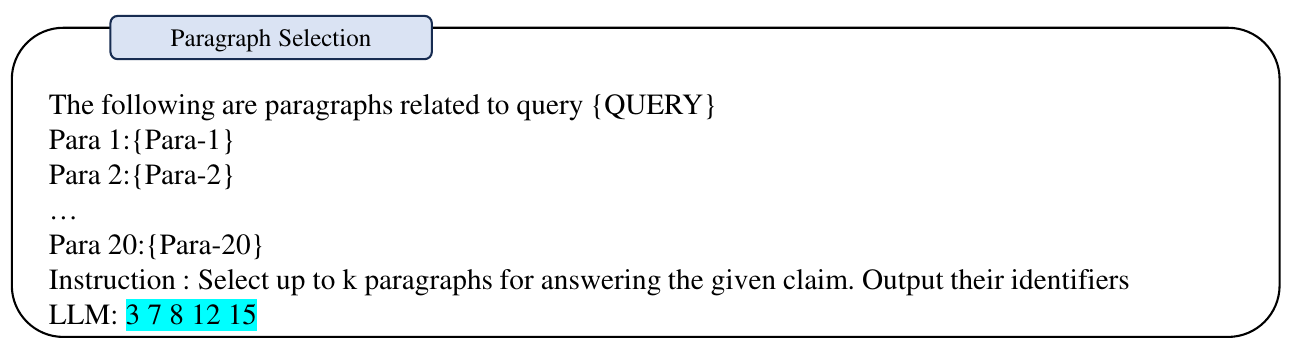}
        \caption{LLM-guided filtering and update of paragraph set \( P \).}
        \label{fig4b}
    \end{subfigure}
    
    \caption{Illustration of progressive paragraph selection during multi-step retrieval.}
    \label{fig4}
\end{figure}

Here, \( R \) denotes the retriever, and \( r(C, p) \) represents the similarity score between the claim and paragraph, computed using a dense retriever \cite{43}. However, dense retrievers \cite{44} are based on dual encoder architectures with limited capacity to capture token-level interactions, which reduces their ability to accurately rank fine-grained semantic relevance \cite{46}.
To address this limitation, we adopt a \textit{progressive selection mechanism} inspired by retrieval-re-ranking frameworks \cite{47}. As shown in Figure ~\ref{fig4}, the LLM incrementally scans candidate paragraphs and selects a refined subset. A sliding window is applied over the current paragraph set \( P \) to extract new candidates \( P_c^\ast \) (Figure~\ref{fig4a}). From the union \( P \cup P_c^\ast \), the LLM selects the top-\(k\) most relevant paragraphs with respect to claim \( C \), resulting in an updated paragraph set \( P \) (Figure~\ref{fig4b}). This process removes irrelevant content, enriches semantic coverage, and preserves the set size for efficient downstream processing. Compared to traditional re-ranking, this method directly outputs an updated paragraph set, reducing redundancy and improving retrieval precision through iterative refinement.

\textbf{Key Evidence Selection}: After paragraph-level filtering, we identify sentence-level evidence via semantic similarity. This step aims to extract the most relevant sentences from the refined paragraph set \( P\). Sentence selection is modelled as a semantic matching task between each sentence and the given claim \( C \). Since the retrieval scope has already been narrowed, an exhaustive search over relevant paragraphs is feasible. For a given claim \( c \), the relevance score \( r(c, s_i) \) is computed for each sentence \( s_i \in P \), and the top-\(j\) sentences are selected to form the final evidence set: $\mathbf{E}_\mathbf{s} = \{\mathbf{E}_1, \mathbf{E}_2, \mathbf{E}_3, \ldots, \mathbf{E}_\mathbf{j}\}$.
This selection relies on the vector representations of both the claim and candidate sentences.

\begin{figure}[h]
\centering
\includegraphics[width=0.95\columnwidth]{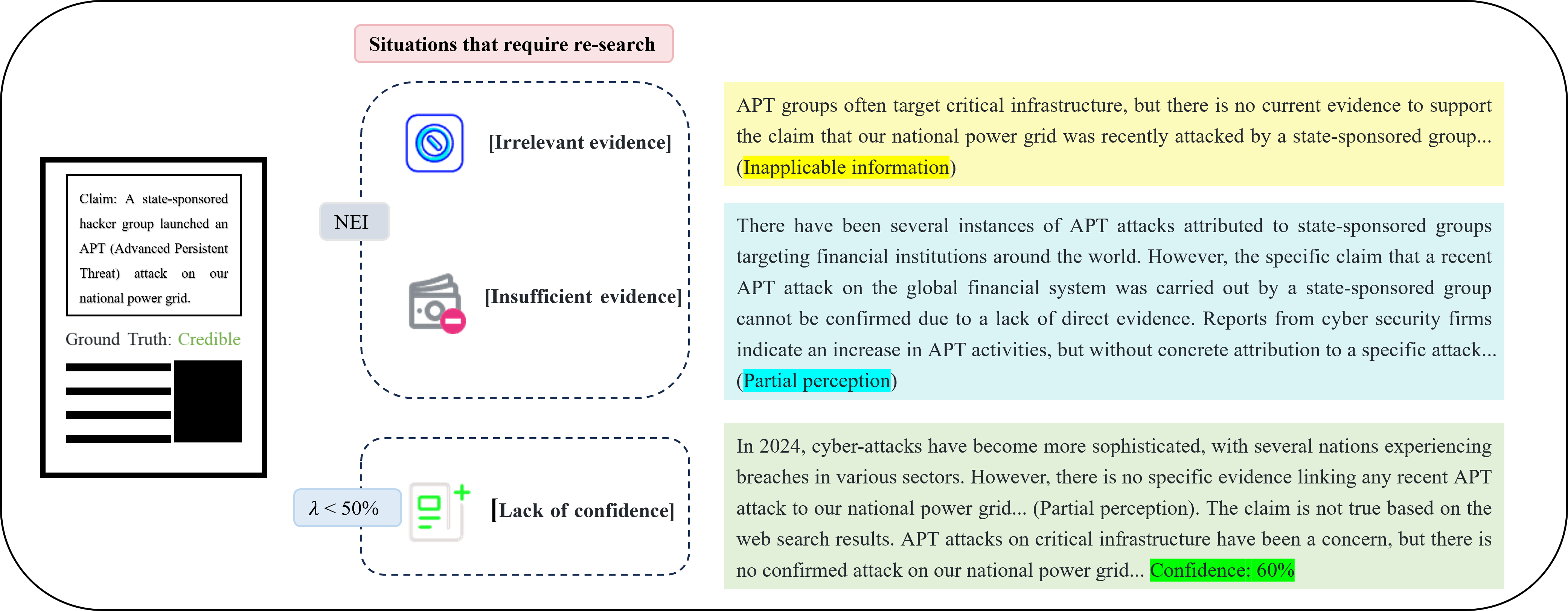}
\caption{
Re-retrieval trigger conditions. \textit{Irrelevant evidence} refers to content unrelated to the claim. \textit{Insufficient evidence} indicates a lack of support for definitive conclusions. \textit{Low confidence} suggests uncertainty in the current judgment, requiring further evidence retrieval.
}
\label{fig5}
\end{figure}

\textbf{Re-retrieval Module}: As shown in Figure~\ref{fig5}, multi-step re-retrieval is triggered under three conditions: irrelevant evidence, insufficient support, or low confidence. This mechanism ensures more comprehensive evidence collection, thus improving the credibility and robustness of the final decision.

When re-retrieval is activated, the model first incorporates the previously gathered evidence into an ``established evidence'' pool. It then generates ``updated claims'' to guide additional retrieval. This iterative loop enables the gradual expansion of the evidence base.

Formally, when the current evidence is insufficient, the LLM outputs the signal ``Not Enough Information (NEI),'' triggering another retrieval cycle:

\begin{equation}
\hat{z}, R_e, \lambda = L(C, E_s, P)
\label{eq4}
\end{equation}

Where \( \hat{z} \in \{\text{Credible, Incredible, NEI}\} \), \( R_e \) is the generated explanation, and \( \lambda \) denotes the confidence score. If \( \hat{z} = \text{NEI} \) or \( \lambda < \theta \), the retrieval loop continues.

\subsection{Text Reasoning Module}

The credibility of a claim is determined by evaluating the semantic alignment between the retrieved evidence and the textual claim. Incorporating explicit evidence in the reasoning process enhances the interpretability and reliability of CTI verification.

The final stage of the LRCTI framework involves verifying the CTI claim using the multi-step retrieved evidence. The selected evidence set \( E \), obtained from the CTI corpus, is aggregated into a prompt and passed to a LLM for reasoning. The LLM evaluates the evidence and determines whether the current information is sufficient or if further evidence is required. The prompt instructs the LLM to classify the claim into one of three categories: \textit{credible}, \textit{incredible}, or \textit{not enough information} (NEI), based on the completeness and consistency of the evidence.

If the classification result is NEI, the model generates an ``established evidence'' set and corresponding ``updated claims'' to initiate additional retrieval cycles. Here, ``established evidence'' refers to previously validated and compressed evidence retained for future reasoning, while ``updated claims'' represent new, potentially simplified sub-claims derived to support further retrieval. The prompts used for this reasoning step are provided in Appendix~\ref{app1}.

To improve consistency and robustness, a confidence score is assigned to each LLM-generated prediction. Both newly retrieved and established evidence are aggregated to support future evaluation cycles. The third decision category includes this accumulated evidence set \( E \), enriched with the previously established evidence pool.

To address issues such as inconsistent answers and hallucinated conclusions, prior work \cite{51} has adopted techniques such as self-consistency and self-judgment, enabling LLM to assign confidence scores within the range of \([0, 100\%]\). However, recent studies \cite{52} have observed that LLM often exhibit overconfidence. To mitigate this, we introduce an \textit{overconfidence factor} \( \theta \in [0,1] \), which adjusts the raw confidence score produced by the LLM. The final confidence score \( \lambda \) is computed as:

\begin{equation}
\lambda = \theta \times \text{Conf}
\label{eq6}
\end{equation}

In Equation~\ref{eq6}, \( \lambda \) denotes the final adjusted confidence score, \( \text{Conf} \) is the initial confidence score output by the LLM, and \( \theta \) represents the overconfidence correction factor. If \( \lambda < 0.5 \), the model automatically initiates the next iteration of retrieval and reasoning.

\subsection{Corpus construction}
The overall corpus construction process, as illustrated in Figure \ref{figurea1}, is divided into three stages: web scraping, data cleaning, and manual verification and annotation.

Design and Implementation of Web Scraping:

\begin{figure}[h]
\centering
\includegraphics[width=0.9\columnwidth]{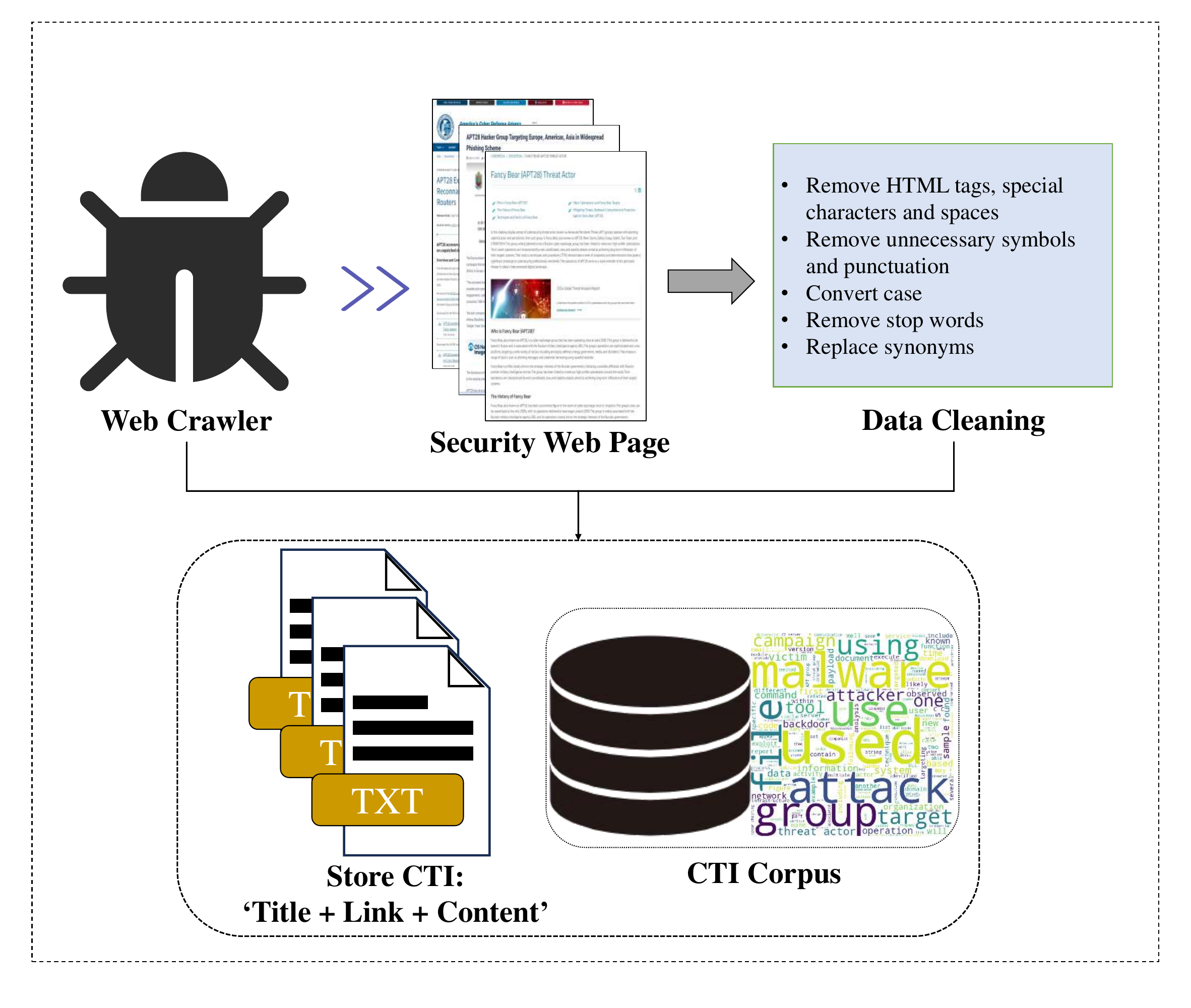} 
\caption{The construction process based on the CTI corpus.}
\label{figurea1}
\end{figure}

We developed a custom web scraping system specifically targeting multiple authoritative data sources in the cybersecurity domain, such as the ATT\&CK framework, Wikipedia, and prominent security blogs. The scraper is capable of identifying newly published content on these platforms and automatically extract relevant information. To handle dynamic and asynchronously loaded content, we employed advanced scraping techniques, such as JavaScript rendering and session management. During the extraction process, we utilized natural language processing (NLP) techniques to identify key threat intelligence elements, such as Indicators of Compromise (IOCs), attack patterns, and descriptions of security incidents.

Data Preprocessing and Formatting:
The raw data extracted often contains a significant amount of non-target information, such as web template elements, navigation links, and advertisements. We implemented a series of text-cleaning steps to remove this extraneous information while using text parsing algorithms to preserve the original structure of the reports, such as headings, paragraphs, and lists. Additionally, we formatted the information into a unified structure to facilitate subsequent indexing and retrieval. This structure typically includes titles, links, and the body content, with each section carefully designed to ensure the completeness and accessibility of the information.

Deduplication and Merging:
In constructing the corpus, we paid particular attention to data quality and consistency. Any duplicate reports were identified and removed using advanced text similarity detection algorithms. Furthermore, we developed a merging algorithm to combine multiple reports about the same security event into a single record, ensuring the completeness of the information. During the merging process, we took special care to retain all critical details, such as timestamps, event descriptions, and related IOCs.

The overall corpus construction process, as illustrated in Figure \ref{figurea1}, is divided into three stages: web scraping, data cleaning, and manual verification and annotation.

Design and Implementation of Web Scraping:
We developed a custom web scraping system specifically targeting multiple authoritative data sources in the cybersecurity domain, such as the ATT\&CK framework, Wikipedia, and prominent security blogs. The scraper is capable of identifying newly published content on these platforms and automatically extracting relevant information. To handle dynamic and asynchronously loaded content, we employed advanced scraping techniques, such as JavaScript rendering and session management. During the extraction process, we utilized natural language processing (NLP) techniques to identify key threat intelligence elements, such as Indicators of Compromise (IOCs), attack patterns, and descriptions of security incidents.

Data Preprocessing and Formatting:
The raw data extracted often contains a significant amount of non-target information, such as web template elements, navigation links, and advertisements. We implemented a series of text-cleaning steps to remove this extraneous information while using text parsing algorithms to preserve the original structure of the reports, such as headings, paragraphs, and lists. Additionally, we formatted the information into a unified structure to facilitate subsequent indexing and retrieval. This structure typically includes titles, links, and the body content, with each section carefully designed to ensure the completeness and accessibility of the information.

Deduplication and Merging:
In constructing the corpus, we paid particular attention to data quality and consistency. Any duplicate reports were identified and removed using advanced text similarity detection algorithms. Furthermore, we developed a merging algorithm to combine multiple reports about the same security event into a single record, ensuring the completeness of the information. During the merging process, we took special care to retain all critical details, such as timestamps, event descriptions, and related IOCs.

\section{Experiments}
To validate the effectiveness of the proposed LRCTI framework, we conducted a comprehensive experimental study designed to address the following research questions:

\begin{itemize}
\item \textbf{RQ1}: Does LRCTI improve the performance of cyber threat intelligence credibility verification and fake news detection compared to existing benchmark models?
\item \textbf{RQ2}: What is the impact of varying the number of retrieval steps in the multi-step retrieval process on overall model performance?
\item \textbf{RQ3}: How does each individual module within LRCTI contribute to enhancing credibility verification effectiveness?
\item \textbf{RQ4}: Are the evidence fragments retrieved through LRCTI’s multi-step retrieval process meaningful and interpretable?
\item \textbf{RQ5}: How does retrieval efficiency differ when using a general-domain corpus (e.g., Wikipedia) versus a domain-specific Cyber Threat Intelligence corpus?
\end{itemize}
\subsection{Experimental Setup}
\subsubsection{Datasets}
\begin{table}[t]
\centering 
\begin{tabular}{ccc} 
\toprule[2pt] 
Platform     & CTI-200 & PolitiFact \\ 
\midrule
\#Credible$|$Real News  & 547 & 399       \\ 
\#Incredible$|$Fake News  & 453 & 345       \\ 
\#Total      & 1000 & 744      \\ 
\bottomrule[2pt]
\end{tabular}
\caption{Statistical data for the two datasets.}
\label{tab1}
\end{table}
To evaluate the effectiveness of the proposed LRCTI framework, we conducted extensive experiments on two real-world datasets: the PolitiFact news dataset and the CTI-200 cyber threat intelligence dataset. The PolitiFact dataset, obtained via FakeNewsNet \cite{53}, comprises news articles labelled as either real or fake. These labels are based on assessments made by professional journalists and fact-checkers who evaluate political news published across a variety of online platforms. The CTI-200 dataset \cite{54} contains cyber threat intelligence reports collected from multiple cybersecurity websites. Each entry is categorized as either credible or incredible based on its source and content reliability. The key statistical characteristics of both datasets are summarized in Table~\ref{tab1}.

\subsubsection{Baseline}
We evaluated the LRCTI model against 10 benchmark methods, divided into two categories: Six classical and recent evidence-based methods(G1), Four LLM-based methods, with or without retrieval components(G2).

\textbf{Evidence-based methods:}
\begin{itemize}
\item DeClarE \cite{6}: Utilizes BiLSTM to embed the semantics of evidence and computes evidence scores through an attentional interaction mechanism.
\item HAN \cite{5}: Employs GRU (Gated Recurrent Unit) embeddings with two modules: one for topic coherence and the other for semantic entailment, facilitating the simulation of claim-evidence interactions.
\item EHIAN \cite{55}: Leverages evidence-aware hierarchical interactive attention networks to explore plausible evidence semantics, enabling interpretable claim verification.
\item MAC \cite{7}: Combines multi-head word-level attention and multi-head document-level attention, providing interpretation for fake news detection at both the word and evidence levels.
\item GET \cite{8}: Models claims and evidence as graph-structured data to explore complex semantic structures, reducing information redundancy through semantic structure refinement layers.
\item MUSER \cite{9}: Implements a multi-step evidence retrieval strategy, exploiting the interdependencies between multiple pieces of evidence to enhance performance.

\end{itemize}

\textbf{LLM-based methods with or without retrieval:}
\begin{itemize}
\item GPT-3.5-turbo  \cite{56}: A sister model to InstructGPT, designed to understand and respond to prompts with detailed explanations. The benchmark model used here is GPT-3.5-turbo.
\item ChatGLM2-6B \cite{57}: An open-source bilingual LLM developed by THUDM, optimized for both Chinese and English conversations. It features improved inference performance and longer context support compared to its predecessor, ChatGLM.
\item WEBGLM \cite{39}: A web-enhanced question-and-answer system based on the Generalized Language Model (GLM), integrating relevant content from the internet and analyzing it through LLMs. The benchmark used here is the 2B version with Bing search integration.
\item ProgramFC \cite{58}: A fact-checking model that decomposes complex claims into simpler sub-statements, which are then solved using a library of specialized functions. It employs strategic retrieval powered by Codex for fact-checking. The benchmark-setting used is their public configuration.
\end{itemize}

\subsubsection{Experimental details}
CTI credibility verification is typically treated as a binary classification problem, evaluated using F1, precision, recall, F1-Macro, and F1-Micro metrics. We use OpenAI's GPT-3.5-Turbo API for retrieval and generation, with hyperparameter  \(\theta \) = 0.7. The LLM is configured with temperature = 0, top-p = 0.75, and a token limit of 4,096. Paragraph retrieval employs a 20-sentence window and selects 50 candidate paragraphs per claim. Baseline hyperparameters follow their original papers, with key parameters optimized for performance.

\begin{table*}[ht]
\centering
\begin{tabular}{@{}lccccccccc@{}}
\toprule[2pt]
\multicolumn{2}{c}{\multirow{2}*{Method}} & \multicolumn{8}{c}{CTI-200} \\
\cmidrule(lr){3-10}
 & & F1-Ma & F1-Mi & F1-T & P-T & R-T & F1-F & P-F & R-F \\ \hline
\multirow{6}{*}{G1} & DeClarE & 0.725 & 0.786 & 0.594 & 0.610 & 0.579 & 0.857 & 0.852 & 0.863 \\ 
& HAN & 0.752 & 0.802 & 0.636 & 0.625 & 0.647 & 0.868 & 0.876 & 0.861 \\ 
& EHIAN & 0.784 & 0.828 & 0.684 & 0.617 & 0.768 & 0.885 & 0.882 & 0.890 \\ 
& MAC & 0.786 & 0.833 & 0.687 & 0.700 & 0.686 & 0.886 & 0.886 & 0.887 \\ 
& GET & 0.800 & 0.846 & 0.705 & 0.721 & 0.694 & 0.895 & 0.890 & 0.902 \\ 
& MUSER & 0.858 & 0.894 & 0.817 & 0.791 & 0.740 & 0.912 & 0.908 & 0.913 \\ \midrule 
\multirow{5}{*}{G2} & GPT-3.5-turbo & 0.782 & 0.798 & 0.762 & 0.783 & 0.752 & 0.802 & 0.787 & 0.807 \\ 
& ChatGLM2-6B & 0.780 & 0.795 & 0.760 & 0.780 & 0.750 & 0.800 & 0.785 & 0.805 \\ 
& WEBGLM-2B  & 0.840 & 0.855 & 0.830 & 0.845 & 0.820 & 0.860 & 0.845 & 0.865 \\ 
& ProgramFC & 0.850 & 0.865 & 0.840 & 0.855 & 0.830 & 0.870 & 0.855 & 0.875 \\
\midrule
{} & \textbf{LRCTI} & \textbf{0.909*} & \textbf{0.936*} & \textbf{0.869*} & \textbf{0.909*} & \textbf{0.833*} & \textbf{0.949*} & \textbf{0.933*} & \textbf{0.965*} \\
\bottomrule[2pt]
\end{tabular}
\caption{Performance comparison on CTI-200}
\label{tab2}
\end{table*}

\begin{table*}[ht]
\centering
\begin{tabular}{@{}lccccccccc@{}}
\toprule[2pt]
\multicolumn{2}{c}{\multirow{2}*{Method}} & \multicolumn{8}{c}{PolitiFact} \\
\cmidrule(lr){3-10}
 & & F1-Ma & F1-Mi & F1-T & P-T & R-T & F1-F & P-F & R-F \\ \hline
\multirow{6}{*}{G1} & DeClarE & 0.654 & 0.651 & 0.656 & 0.689 & 0.673 & 0.651 & 0.613 & 0.664 \\ 
& HAN & 0.661 & 0.660 & 0.679 & 0.676 & 0.682 & 0.643 & 0.650 & 0.637 \\ 
& EHIAN & 0.664 & 0.663 & 0.674 & 0.680 & 0.651 & 0.650 & 0.628 & 0.627 \\ 
& MAC & 0.678 & 0.675 & 0.700 & 0.695 & 0.704 & 0.653 & 0.655 & 0.645 \\ 
& GET & 0.694 & 0.692 & 0.725 & 0.712 & 0.770 & 0.669 & 0.720 & 0.665 \\ 
& MUSER & 0.732 & 0.729 & 0.757 & 0.735 & 0.780 & 0.702 & 0.728 & 0.681 \\ \midrule 
\multirow{4}{*}{G2} & GPT-3.5-turbo & 0.567 & 0.553 & 0.570 & 0.557 & 0.561 & 0.559 & 0.562 & 0.573 \\ 
& ChatGLM2-6B & 0.522 & 0.515 & 0.529 & 0.531 & 0.526 & 0.518 & 0.520 & 0.519 \\ 
& WEBGLM-2B  & 0.628 & 0.633 & 0.601 & 0.617 & 0.639 & 0.612 & 0.660 & 0.626 \\ 
& ProgramFC & 0.684 & 0.678 & 0.733 & 0.725 & 0.741 & 0.635 & 0.622 & 0.643 \\
\midrule
 & \textbf{LRCTI} & \textbf{0.861*} & \textbf{0.861*} & \textbf{0.883*} & \textbf{0.899*} & \textbf{0.870*} & \textbf{0.838*} & \textbf{0.842*} & \textbf{0.835*} \\
\bottomrule[2pt]
\end{tabular}
\caption{Performance comparison on PolitiFact}
\label{tab3}
\end{table*}

\subsection{Performance results (RQ1)}

LRCTI was evaluated against ten baseline models, including six evidence-based methods and four LLM-based approaches. As shown in Tables~\ref{tab2} and~\ref{tab3}, LRCTI consistently outperformed state-of-the-art baselines across both real-world datasets, achieving relative improvements of 5\% and 10\% in F1-Macro and F1-Micro scores on the CTI-200 and PolitiFact datasets, respectively.

To further assess robustness, we conducted evaluations using both ``credible-as-positive'' and ``incredible-as-positive'' settings. Under both configurations, LRCTI achieved superior performance in terms of F1 score, precision, and recall. Since precision is numerically equivalent to F1-Macro in our experimental setting, it was excluded from the final aggregated evaluation for clarity.

In addition to its overall classification accuracy, LRCTI demonstrated strong performance in both CTI credibility verification and fake news detection tasks. On the CTI-200 dataset, it improved F1-False, Precision-False, and Recall-False by 3\%, 2.5\%, and 5\%, respectively, with comparable gains observed on the PolitiFact dataset. These results highlight the effectiveness of LRCTI's LLM-driven multi-step evidence retrieval framework, which enhances the relevance and quality of retrieved evidence for veracity verification.

\begin{figure}[t]
\centering
\includegraphics[width=0.95\columnwidth]{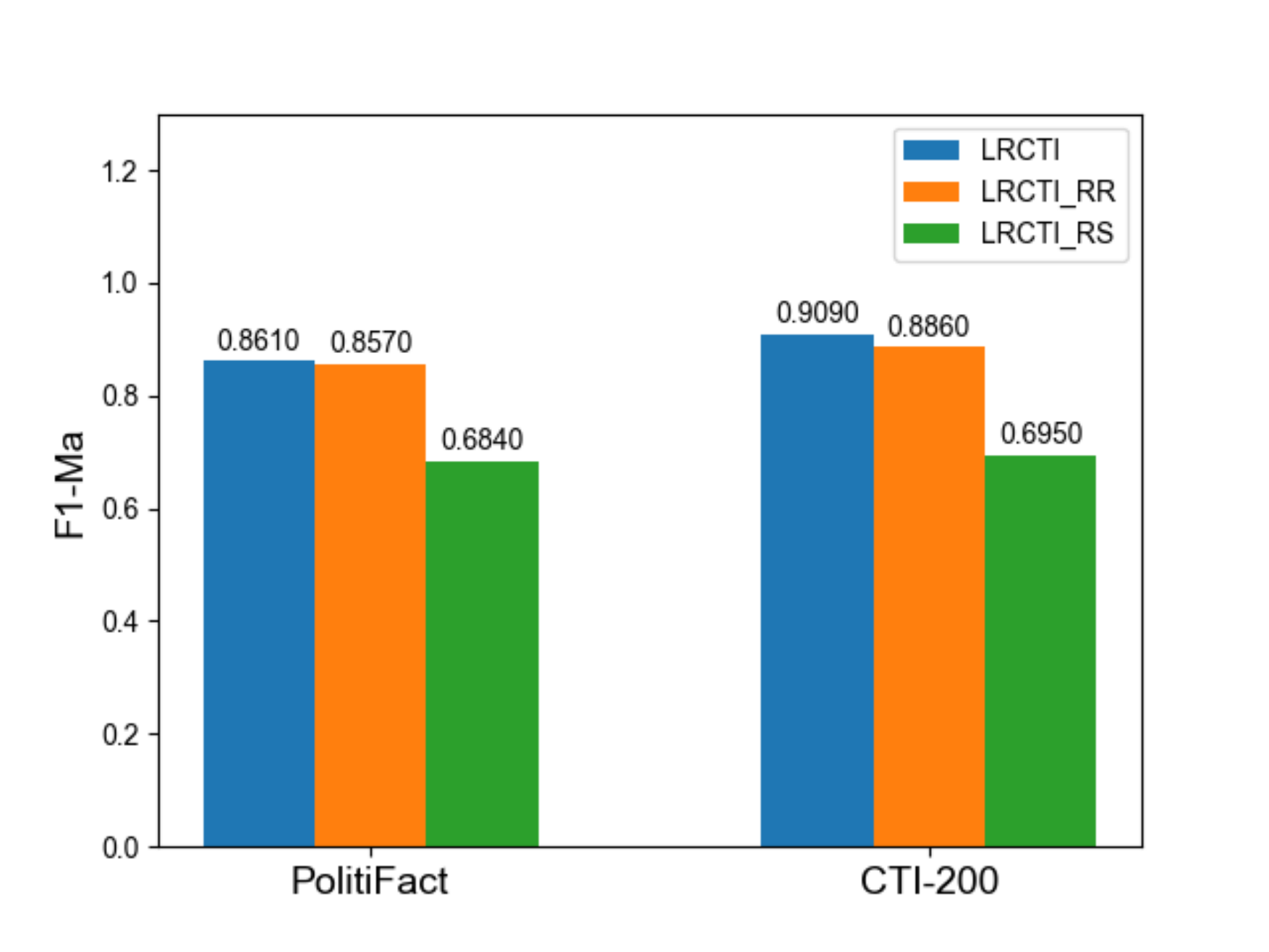} 
\caption{Ablation Study Results: LRCTI represents the performance of the complete model, LRCTI-RR indicates the model with the multi-step retrieval module removed, and LRCTI-RS represents the model with the text summarization module removed.}
\label{fig6}
\end{figure}

\begin{figure}[t]
\centering
\includegraphics[width=0.95\columnwidth]{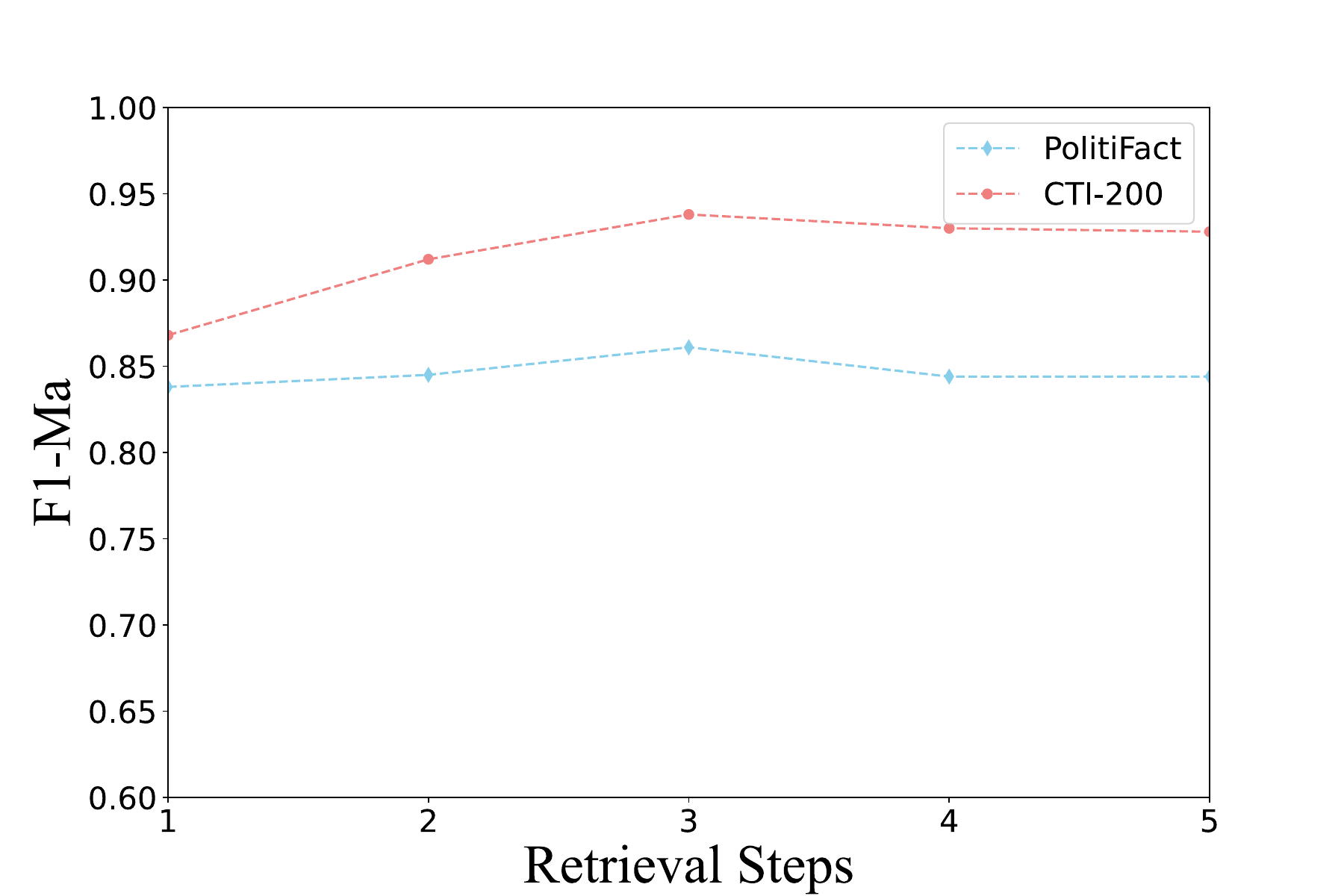} 
\caption{Ablation experiments with different numbers of retrievals for F1-Ma performance, blue line on PolitiFact news dataset, red line on CTI-200 dataset.}
\label{fig7}
\end{figure}
\subsection{Ablation studies (RQ 2, RQ3)}
In this section, we conduct a series of comparative experiments to evaluate the contribution of each module within the LRCTI framework. As illustrated in Figure~\ref{fig6}, the full LRCTI model significantly outperforms its variant LRCTI-RR, where the multi-step re-retrieval component is removed. This result highlights the critical role of iterative evidence refinement in enhancing model performance.

Furthermore, the text summarization module proves essential for filtering out irrelevant information and isolating key CTI claims. Its removal leads to a noticeable drop in prediction accuracy, demonstrating its importance in denoising and improving input quality. LRCTI also outperforms both LRCTI-RS (without summarization) and LRCTI-RR (without re-retrieval), confirming that the elimination of either component results in performance degradation. These findings validate the effectiveness of each primary module in the overall framework.

We further investigate the impact of varying the number of retrieval steps in the multi-step retrieval module. For this analysis, we evaluate model performance on both datasets using the F1-Macro score, as shown in Figure~\ref{fig7}. Results indicate that performance consistently improves as the number of retrieval steps increases. This improvement stems from the model’s ability to refine its search iteratively when the initial evidence retrieval fails to yield sufficiently relevant information.

However, beyond three retrieval steps, performance gains plateau and may even degrade slightly, possibly due to the inclusion of redundant or less relevant content. Notably, the optimal number of retrieval steps—three—remains consistent across datasets of varying complexity, suggesting that LRCTI’s iterative mechanism generalizes well to different CTI and fact-checking scenarios.

\subsection{Interpretability studies (RQ 4)}

\begin{figure}[h]
\centering
\includegraphics[width=1.0\columnwidth]{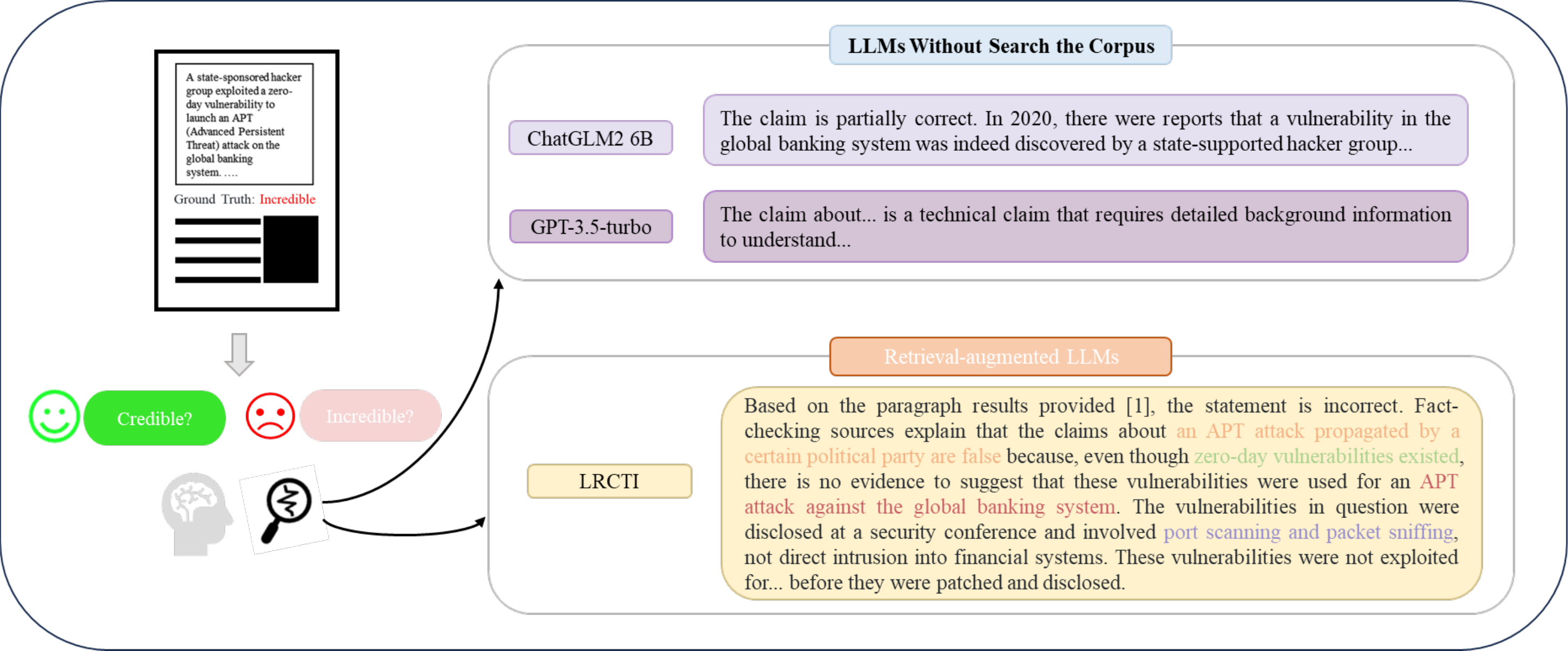} 
\caption{Explanation of the Case Study: Text with a colored background indicates the cited evidence.}
\label{fig8}
\end{figure}

\textbf{Case Study} \quad In this section, we present a qualitative case study to demonstrate the explanatory capabilities of the LRCTI framework. As shown in Figure~\ref{fig8}, we analyze a CTI claim: \textit{``A state-sponsored hacker group exploited a zero-day vulnerability to launch an APT (Advanced Persistent Threat) attack on the global banking system.''} Through the processes of multi-step evidence retrieval and structured reasoning, LRCTI correctly classifies this claim as \textit{incredible}. The model effectively aggregates and synthesizes key evidence fragments and organizes them into a coherent, human-readable explanation. Notably, LRCTI demonstrates strong capabilities in structuring its reasoning in an interpretable manner, clearly distinguishing between retrieved factual information and LLM-generated inferences. This attribution mechanism enhances transparency and supports user trust in the system's outputs. Overall, the case illustrates the model's strength in producing not only accurate predictions but also interpretable justifications that aid human understanding.

\begin{table}[h]
    \centering
    \begin{tabular}{|c|c|c|}
    \hline
        \textbf{Corpus} & \textbf{Efficiency} & \textbf{r(c,e)} \\ \hline
        \textbf{Wiki} & 10s & 0.733  \\ \hline
        \textbf{CTI} & \textbf{5s} & \textbf{0.895}  \\ \hline
    \end{tabular}
    \caption{The comparison of retrieval efficiency of the model between the Wikipedia corpus and the CTI corpus.}
       \label{tab6}
\end{table}

\begin{table}[h]
    \centering
    \begin{tabular}{|c|c|c|c|}
    \hline
        \textbf{Method} & \textbf{F1} & \textbf{Precision} & \textbf{Agreement } \\ \hline
        \textbf{MUSER} & 0.758 & 0.733 & 76.7\%  \\ \hline
        \textbf{LRCTI} & \textbf{0.815} & \textbf{0.812} & \textbf{81.6\%}  \\ \hline
    \end{tabular}
    \caption{Results of the user study. The agreement measure
means the proportion of concurrence between the user's
judgment and the model's judgment.}
    \label{tab5}
\end{table}

\textbf{User Study} \quad To evaluate the practical effectiveness of LRCTI in assisting human judgment, we conducted a user study to verify whether real-world users can accurately evaluate the credibility of CTI claims based on retrieved evidence. A total of 40 claims were randomly selected from the CTI-200 dataset, including 20 labelled as credible and 20 as incredible. We compared the quality of supporting evidence provided by LRCTI against that of the MUSER model. Eight university students participated in the evaluation, each independently reviewing a randomized subset of claim–evidence pairs without communication or collaboration. For each pair, participants were given 3 minutes to read the retrieved evidence and make a binary credibility judgment (credible/incredible). Additionally, they rated their confidence in each decision on a 5-point Likert scale. The evidence presented to participants was sourced exclusively from either LRCTI or MUSER, with each claim associated with only one model to avoid bias. The results, summarized in Table~\ref{tab5}, indicate that LRCTI significantly outperforms MUSER in terms of evidence quality. Participants demonstrated higher accuracy and confidence when verifying claims using LRCTI-generated evidence, highlighting its effectiveness in supporting human interpretability and decision-making.

\subsection{Comparison of corpus retrieval efficiency (RQ 5)}

In this experiment, we compared the retrieval efficiency of the Wikipedia corpus and the Cyber Threat Intelligence (CTI) corpus using a common set of claims derived from CTI reports. For consistency, the evidence retrieval process was conducted with a fixed correlation threshold of \(\boldsymbol{\lambda} = 0.5\).

As shown in Table~\ref{tab6}, relevant evidence was retrieved from the CTI corpus in an average of 5 seconds per query, whereas retrieval from the Wikipedia corpus required an average of 10 seconds. In addition to speed, the semantic relevance of retrieved evidence, measured by the similarity score \(\boldsymbol{r(c, e)}\), was also consistently higher for the CTI corpus.

These results demonstrate that the CTI corpus provides both higher retrieval efficiency and greater contextual relevance compared to the Wikipedia corpus. The domain specificity of the CTI corpus enables faster access to actionable intelligence, allowing analysts to respond more rapidly and effectively to emerging threats. This improved responsiveness is particularly valuable in time-sensitive cyber threat environments.


\section{Conclusion}
In this paper, we proposed \textbf{LRCTI}, a novel framework for \textit{CTI credibility verification}, enhanced by LLM and a multi-step evidence retrieval mechanism. LRCTI consists of three core components: a summarization module for extracting concise and verifiable threat claims, a multi-step retrieval module for refining and correlating relevant evidence, and a text reasoning module that leverages LLM to perform claim verification and generate interpretable justifications. By systematically aligning CTI claims with retrieved evidence through iterative refinement, LRCTI enhances both the accuracy and explainability of credibility verification. Extensive experiments on two real-world datasets—CTI-200 and PolitiFact—demonstrate that our framework significantly outperforms existing baselines in both credibility verification and fake news detection tasks. Overall, our results validate the effectiveness of LLM-based multi-step reasoning in improving CTI credibility verification. LRCTI enables more transparent, evidence-driven decision-making, thereby supporting analysts in efficiently evaluating the trustworthiness of threat intelligence.

\bibliography{reference}{}

\appendix
\section {Appendices}\label{app1}
\subsection{Prompts’ Description}
\begin{figure*}[ht]
\centering
\includegraphics[width=0.9\textwidth]{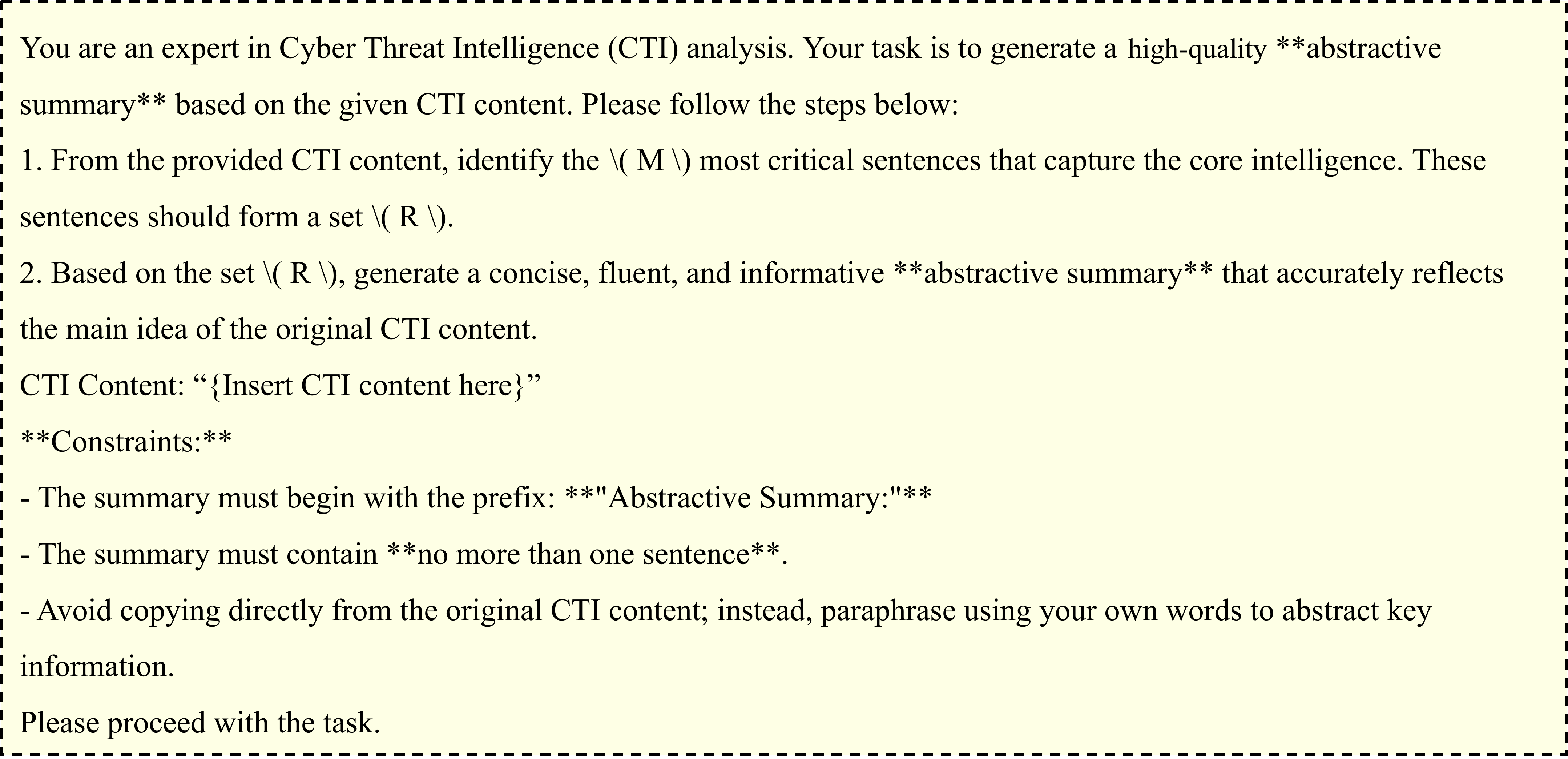} 
\caption{ Text summary(ts) Prompt}
\label{figa2}
\end{figure*}

\begin{figure*}[ht]
\centering
\includegraphics[width=0.9\textwidth]{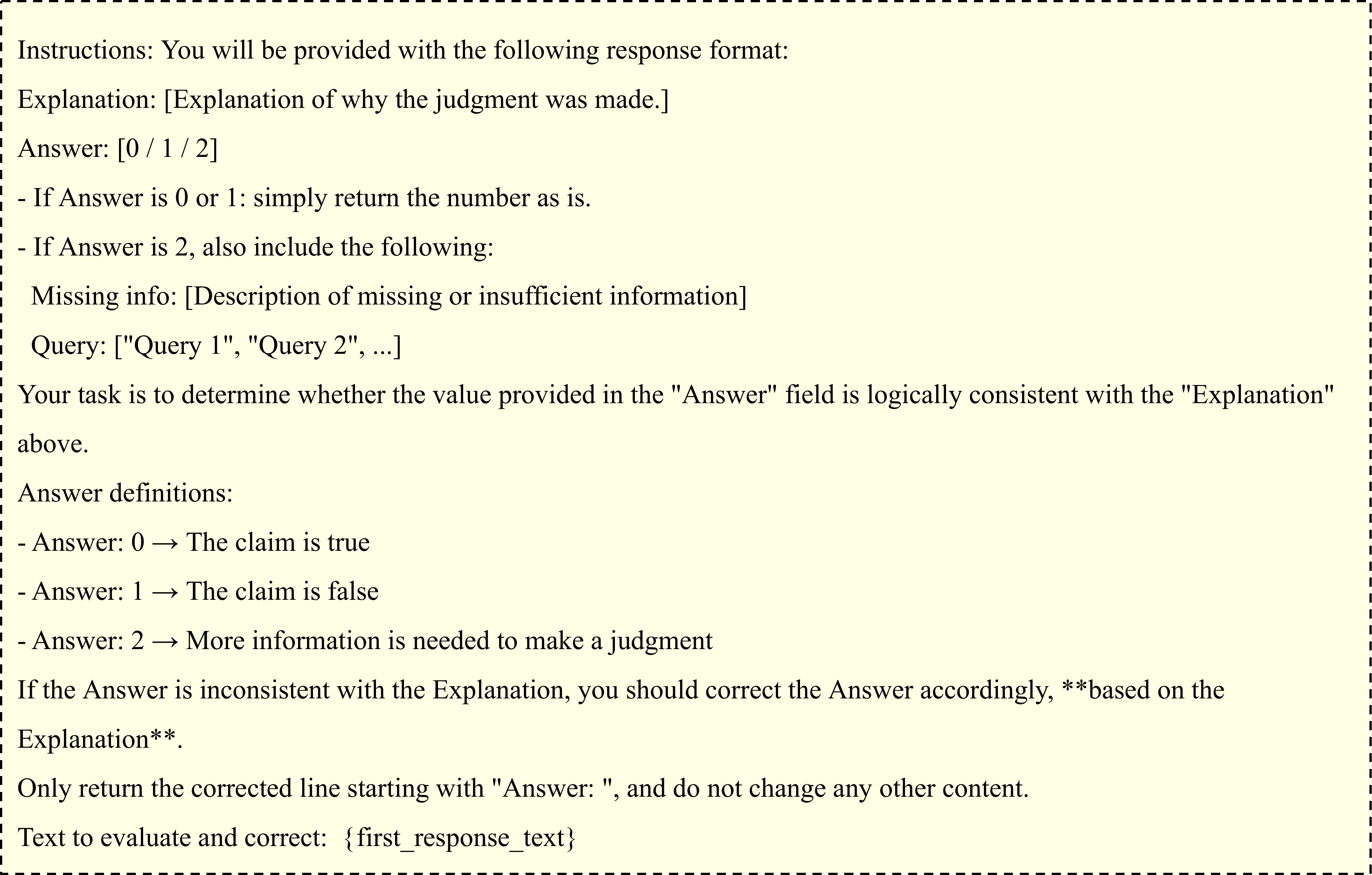} 
\caption{Response correction(rc) Prompt.}
\label{figa3}
\end{figure*}

\begin{figure*}[ht]
\centering
\includegraphics[width=1.0\textwidth]{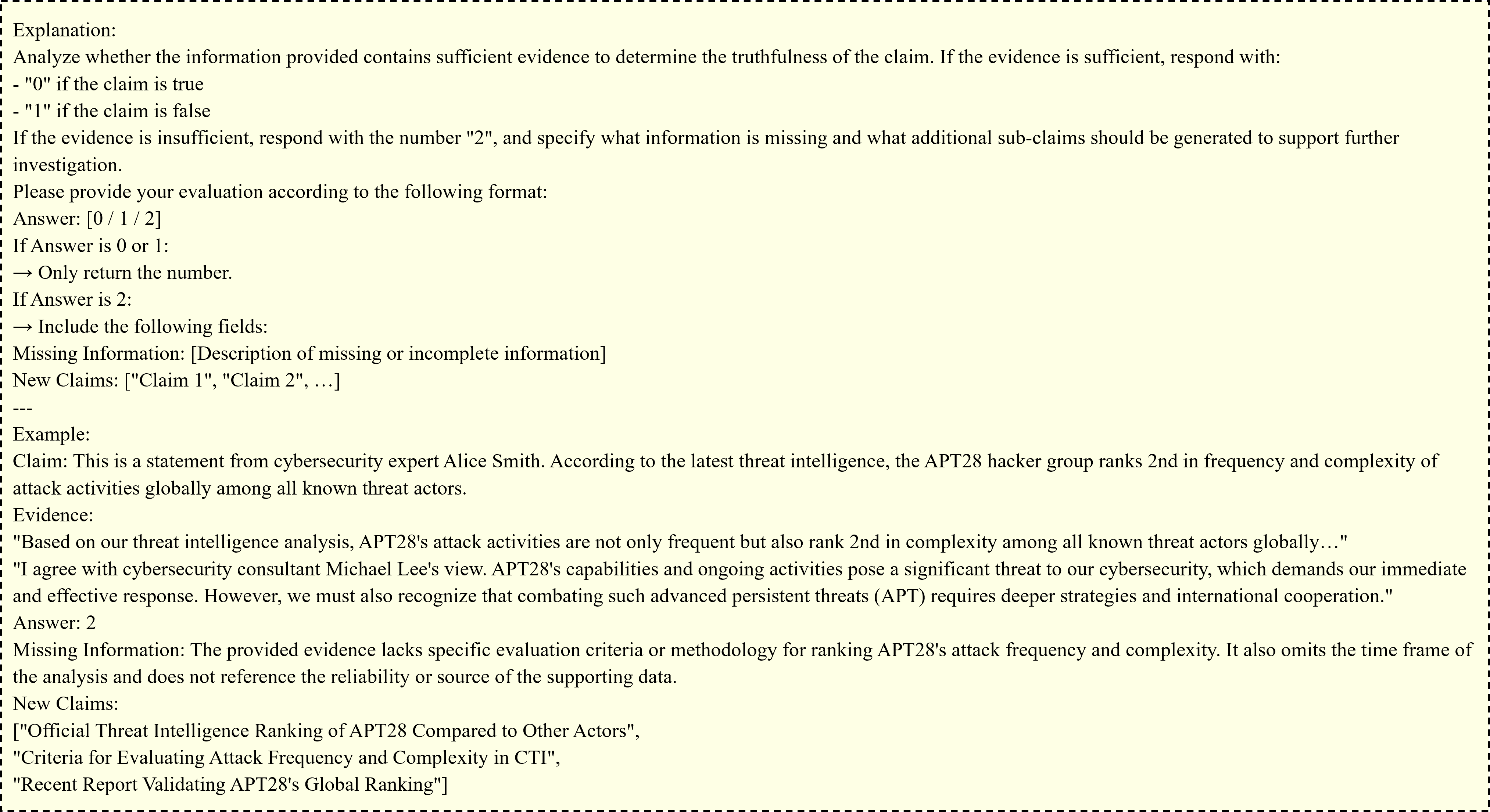} 
\caption{ Step check(sc) Prompt.}
\label{figa4}
\end{figure*}

\textbf{Type Description}
\begin{itemize}
\item Text summary: This prompt is used to convert long news into verifiable short claims, as shown in Figure \ref{figa2}.
\item Response correction: This prompt aims to verify the consistency of LLMs' complete output and address the issue of self-contradictory answers produced by these models, as shown in Figure \ref{figa3}.

\item  Step check: This prompt is used as part of our LRCTI model, as shown in Figure \ref{figa4}.
\end{itemize}


\end{document}